\documentclass{raa}

\usepackage{graphicx,times}             

\begin{document}

  \title{LAMOST Spectral Survey}

\subtitle{An Overview}

  \volnopage{Vol.0 (200x) No.0, 000--000}      
  \setcounter{page}{1}          

  \author{Gang Zhao
     \inst{1}
   \and
     Yongheng Zhao
      \inst{1}
      \and
      Yaoquan Chu
      \inst{1,2}
      \and
      Yipeng Jing
      \inst{3}
  \and 
     Licai Deng
     \inst{1,*}
}

  \institute{Key Laboratory for Optical Astronomy, National Astronomical Observatories, Chinese Academy of Sciences,
            Beijing 100012, China, $^*${\it licai@bao.ac.cn}\\
       \and
             University of Science and Technology of China, No.96, JinZhai Road Baohe District, Hefei, Anhui 230026, China\\
       \and
            Shanghai Astronomical Observatory, CAS, 80 Nandan Road, Shanghai 200030, China\
  }

  \date{Received~~2009 month day; accepted~~2009~~month day}

\abstract{ LAMOST (Large sky Area Multi-Object fiber Spectroscopic Telescope) is a Chinese national scientific research facility operated by National Astronomical Observatories, Chinese Academy of Sciences (NAOC). After two years of commissioning beginning in 2009, the telescope, instruments, software systems and operations are nearly ready to begin the main science survey. Through a spectral survey of millions of objects in much of the northern sky, LAMOST will enable research in a number of contemporary cutting edge topics in astrophysics, such as: discovery of the first generation stars in the Galaxy, pinning down the formation and evolution history of galaxies especially the Milky Way and its central massive black hole, looking for signatures of dark matter distribution and possible sub-structures in the Milky Way halo. To maximize the scientific potential of the facility, wide national participation and international collaboration has been emphasized. The survey has two major components: the LAMOST ExtraGAlactic Survey (LEGAS), and the LAMOST Experiment for Galactic Understanding and Exploration (LEGUE). 
Until LAMOST reaches its full capability, the LEGUE portion of the survey will use the available observing time, starting in 2012. 
An overview of the LAMOST project and the survey that will be carried out in next five to six years is presented in this paper. 
The science plan for the whole LEGUE survey, instrumental specifications, site conditions, the descriptions of the current on-going pilot survey, including its footprints and target selection algorithm, will be presented as separate papers in this volume.
}

  \authorrunning{G. Zhao et al. }            
  \titlerunning{LAMOST spectral survey overview }  

  \maketitle

%
%
\section{Introduction}           
\label{sect:intro}

The Chinese national scientific research facility, LAMOST (Large sky Area Multi-Object fiber Spectroscopic Telescope, also named Guo Shou Jng Telescope (GSJT); for the original project plan see Wang et al. 1996, see also Cui et al . 2010 and Cui, Zhao \& Chu et al. 2012 for more details on the optical system) will begin carrying out its scientific survey of over 10 million stars and galaxies in Oct 2012. The survey contains two main parts: the LAMOST ExtraGAlactic Survey (LEGAS), and the LAMOST Experiment for Galactic Understanding and Exploration (LEGUE) survey of Milky Way stellar structure. Because of the special horizontal reflecting Schmidt design of the optics, the Guoshoujing telescope has a field of view as large as 20 square degrees, and at the same time a large effective aperture that varies from 3.6 to 4.9 meters in diameter (depending on the direction it is pointing). The unique design of LAMOST enables it to take 4000 spectra in a single exposure to a limiting magnitude as faint as $r=19$ at resolution $R=1800$, which is equivalent to the design aim of $r=20$ for the resolution R=500. This telescope therefore has a great potential to efficiently survey a large volume of space for stars and galaxies. 

\begin{figure}
\centerline{\includegraphics[width=10cm]{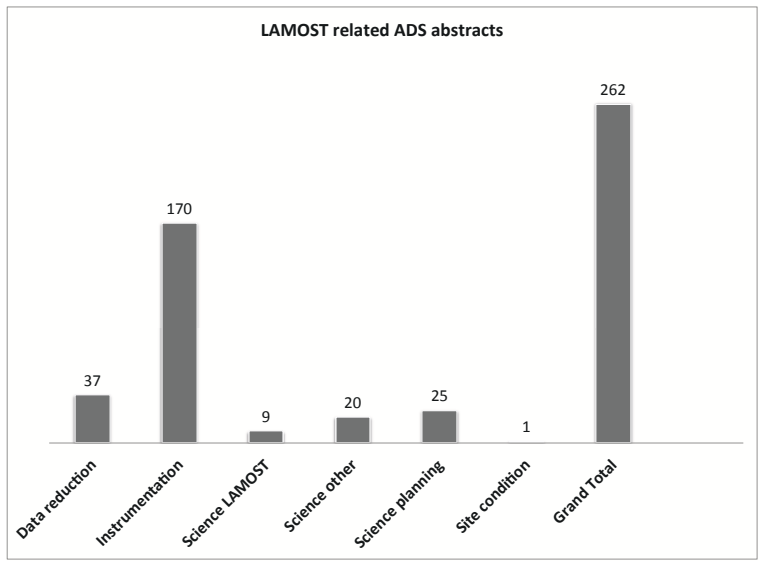}}
\caption{A statistical survey of all the abstracts that are linked to LAMOST from ADS, as of March 2012. }\label{lamost-ads}
\end{figure}

Much progress has been made in the past two years in tuning the performance of LAMOST systems, including fiber positioning, dome seeing control, and optical alignments of spectrographs on hardware side; and calibrations, data pipelines and data archiving on software side.  As demonstrated by 2 years of technical commissioning, the system is not yet reaching its original designed performance. However, LAMOST is already producing useful spectra of bright objects, and can successfully observe targets as faint as $r=20$ in the very best cases (and even fainter for emission line objects). LAMOST commissioning data have already produced a number of scientific results, including a search for metal poor stars (Li et al. 2010), the discovery of new quasars (He et al. 2010, Wu et al. 2010a,b) and planetary nebulae (Yuan et al. 2010),
and mapping of the 2D stellar population pattern of the M31 disk (Zou et al. 2011). A summary of statistics of abstracts from ADS (publications and conference proceedings) on LAMOST scientific work and technical reports is shown in Fig.~\ref{lamost-ads}. Since 1996, over 250 abstracts related to LAMOST have been contributed to the literature, according to the ADS search engine. Most of the abstracts are presentations and papers on instrumentation (65\%) and data reduction algorithms (14\%) . Science planning started at a later time (mostly after 2006) after the detailed hardware designs of the LAMOST project had been determined. 
There are also scientific research presentations and papers that use LAMOST data (3\%) or are somehow related with LAMOST (8\%). All these publications demonstrate the complex technical design of LAMOST and its scientific potential. More recent research papers show that LAMOST not only realized a high efficiency in spectral data acquisition, but also has been capable of producing science quality data.

\section{The LAMOST Project Technical Description}\label{sect:the workhorse}

The novel design of the LAMOST system provides a combination of a large-aperture, large field of view telescope feeding a highly-multiplexed spectroscopic system. However, these capabilities also impose some significant observing constraints
that must be carefully considered when designing the survey for the Milky Way Galaxy and extra-galactic objects.  These constraints are outlined below.

\subsection{General layout}

LAMOST is a reflecting Schmidt telescope with its optical axis fixed along the north-south meridian (Su et al. 1998, Cui et al. 2010). Both the Schmidt plate and the primary mirror are segmented. The Schmidt plate (``Mirror A'') is made of 24 sub-mirrors, and the primary (``Mirror B'') of 37 segments, each of which is in a hexagon shape 1.1 meters in diagonal dimension.  Both mirrors are controlled by active optics. Mirror A is made of thin flat glass that can be deformed to a Schmidt plate surface in real time (Su \& Cui 2004). 

The focal surface is circular with a diameter of 1.75 meters ($\sim5^\circ$); 4000 fibers are almost evenly distributed on it. Each of the fibers can be moved in two degrees of freedom by 2 motors. A Shack-Hartmann system is located in the center of the field of view to enable active mirror deformation. Guiding is provided by 4 guide CCD cameras placed about halfway out from the center of the field; these cameras monitor shifts in the positions of relatively bright stars during the exposures.
The guide stars are also used to measure seeing (based on the FWHM of stars on the images) during the exposure for each plate. 

\begin{figure}
\centerline{\includegraphics[width=7cm]{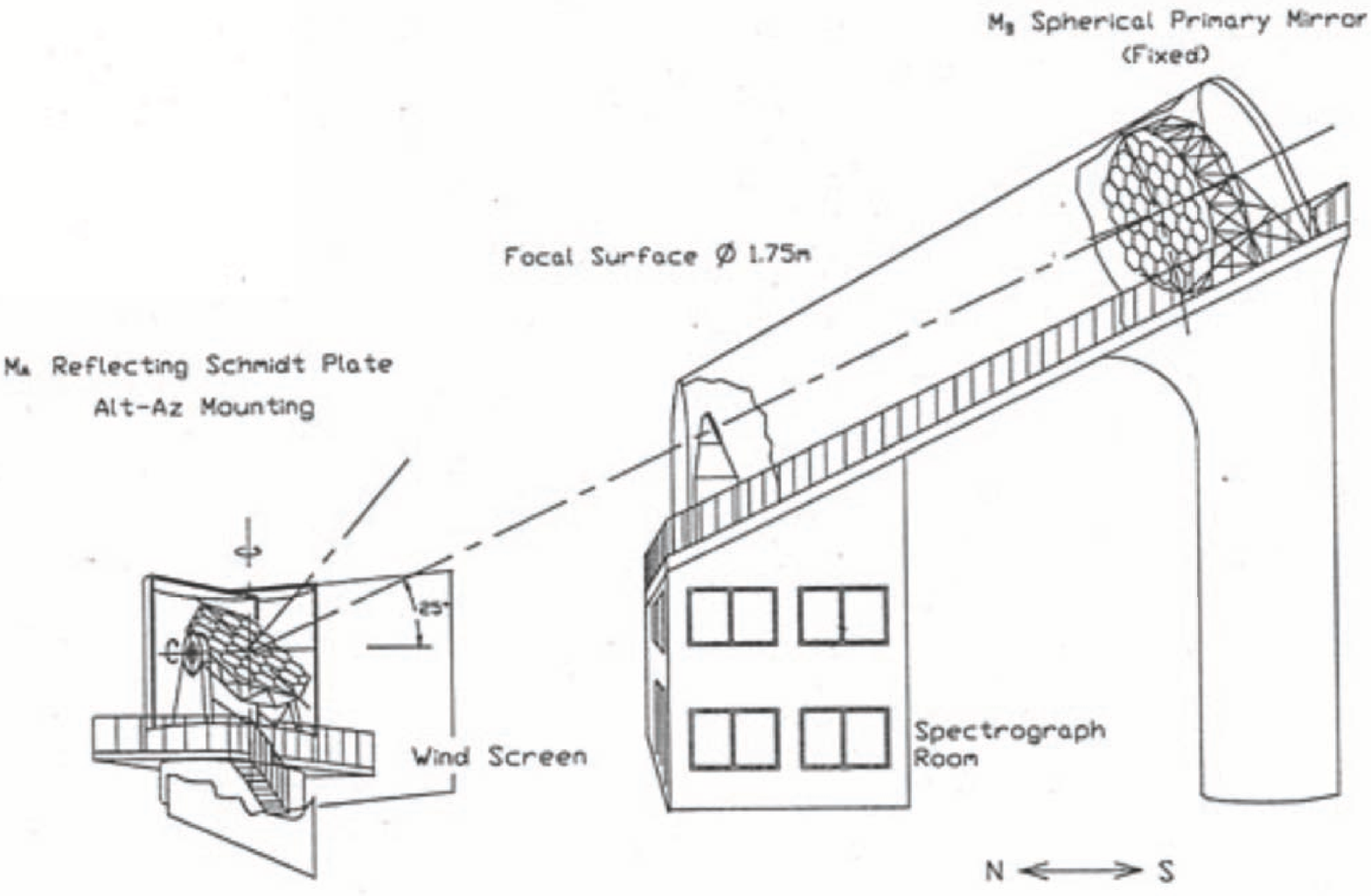}\includegraphics[width=7cm]{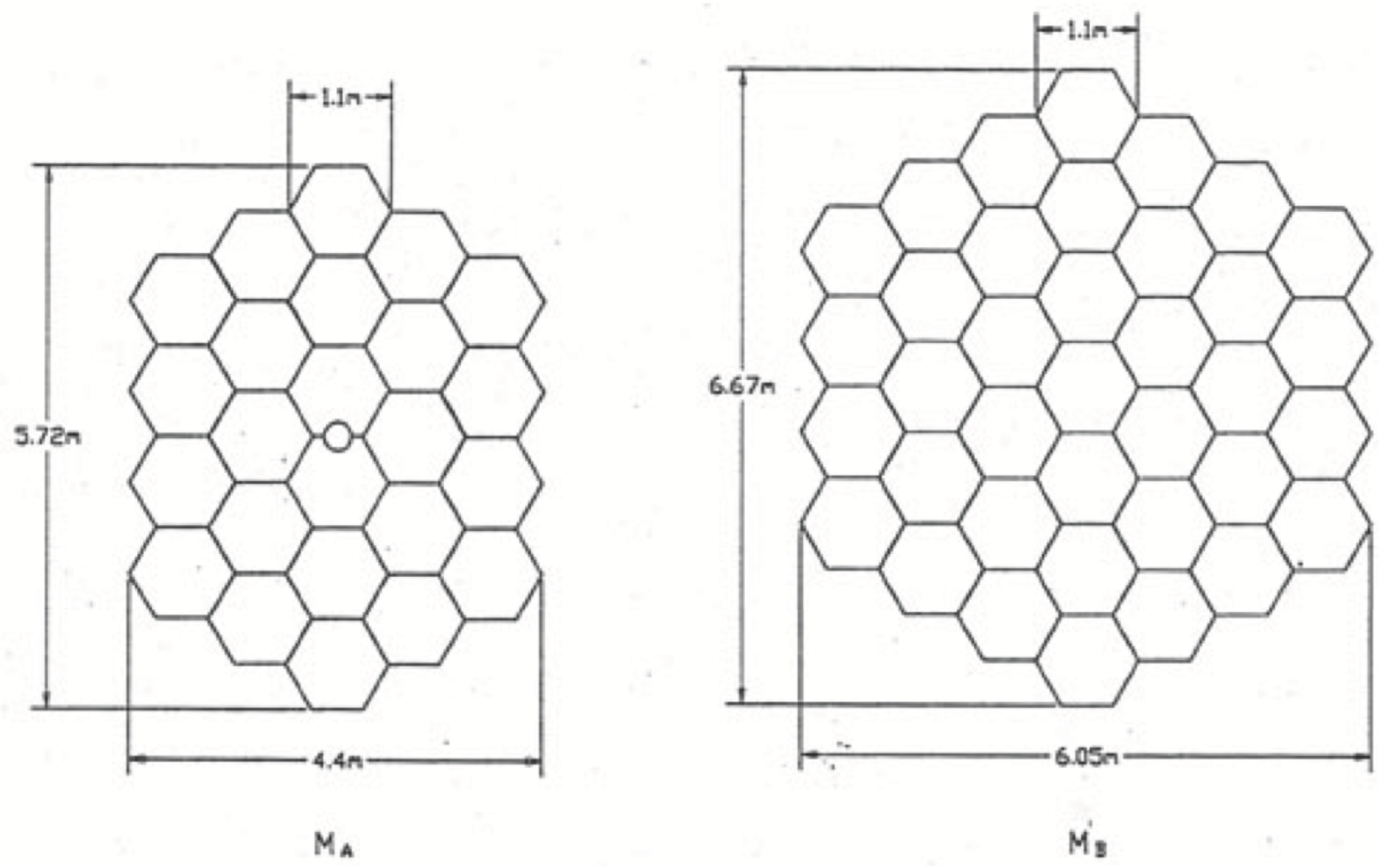}}
\caption{The left panel shows a layout of LAMOST system and light path. The right panel shows the sketch of the two segmented mirrors, the left one is the Schmidt plate with AO, the right one is the spherical primary. The scales are indicated (courtesy Shou-guan Wang, Ding-qiang Su, Yao-quan Chu, Xiangqun Cui and Ya-nan Wang, in LAMOST documentation archive. ).}\label{layout}
\end{figure}

The dome for LAMOST has 3 towers aligned in a North-South direction, hosting Mirror A, the focal surface and instruments, and Mirror B respectively. Mirror A has a transitional dome which can be opened completely. The focal surface and Mirror B towers are connected by a tube structure. The layout of the system and a sketch of the optical path is shown in Figure~\ref{layout}.

\subsection{Pointing} The telescope can be pointed in declination from $-10^\circ$ to
$90^\circ$.  However, for $\delta>60^\circ$, the effective collecting area of the telescope decreases, reducing the depth to which spectra can be obtained.  By
$\delta=90^\circ$, $30\%$ of the light in the center of the field is lost
and the image quality in the outer parts of the field is degraded such that one might
not be able to use the fibers that are more than 1.5 degrees from the center of the
focal surface (eliminating 64\% of the science fibers).  At all declinations, $20\%$ of
the light is lost at the edges of the field of view due to vignetting.  For this reason,
our survey footprint is limited to $\delta<60^\circ$.

The telescope can be pointed about 2 hours ($30^\circ$) in each direction from the meridian,
though the exact limits need to be verified by direct observations.
Since we typically need 1.5 hours of exposure time for each pointing, there are very strong constraints on the right ascension that can be targeted at any given time.  When planning the survey, we must ensure that there are always fields available at each right ascension that becomes available.  It is beneficial to have more targets available than can be observed in the length of the survey, since we will not be able to plan for the exact number of hours available at each right ascension due to the weather.

\subsection{Weather} The weather at the site is extremely seasonal; for a detailed discussion of the site conditions, please refer to Yao et al. (2012) in this volume. The weather statistics by day of the year (Figure 1 in Yao et al. 2012) were taken from observing logs recorded in past years from the BATC (Beijing-Arizona-Taipei-Connecticut multi-color photometric survey) experiment, which operates at the same site.  The figure shows that observing conditions in June, July, and August will give us very little opportunity to observe at right ascensions of 16h to 22h.  The part of the sky for which we will have the greatest number of observations is 2h to 8h, when fields near the Galactic anticenter are available.  We are planning a survey footprint that concentrates the spectroscopic targets in the areas of the sky when we will have more observing time (Deng et al. 2012, Yang et al. 2012, Zhang et al. 2012, Carlin et al. 2012 in this volume).

\subsection{Fiber Spacing}  
\begin{figure}
\centerline{\includegraphics[height=5.5cm]{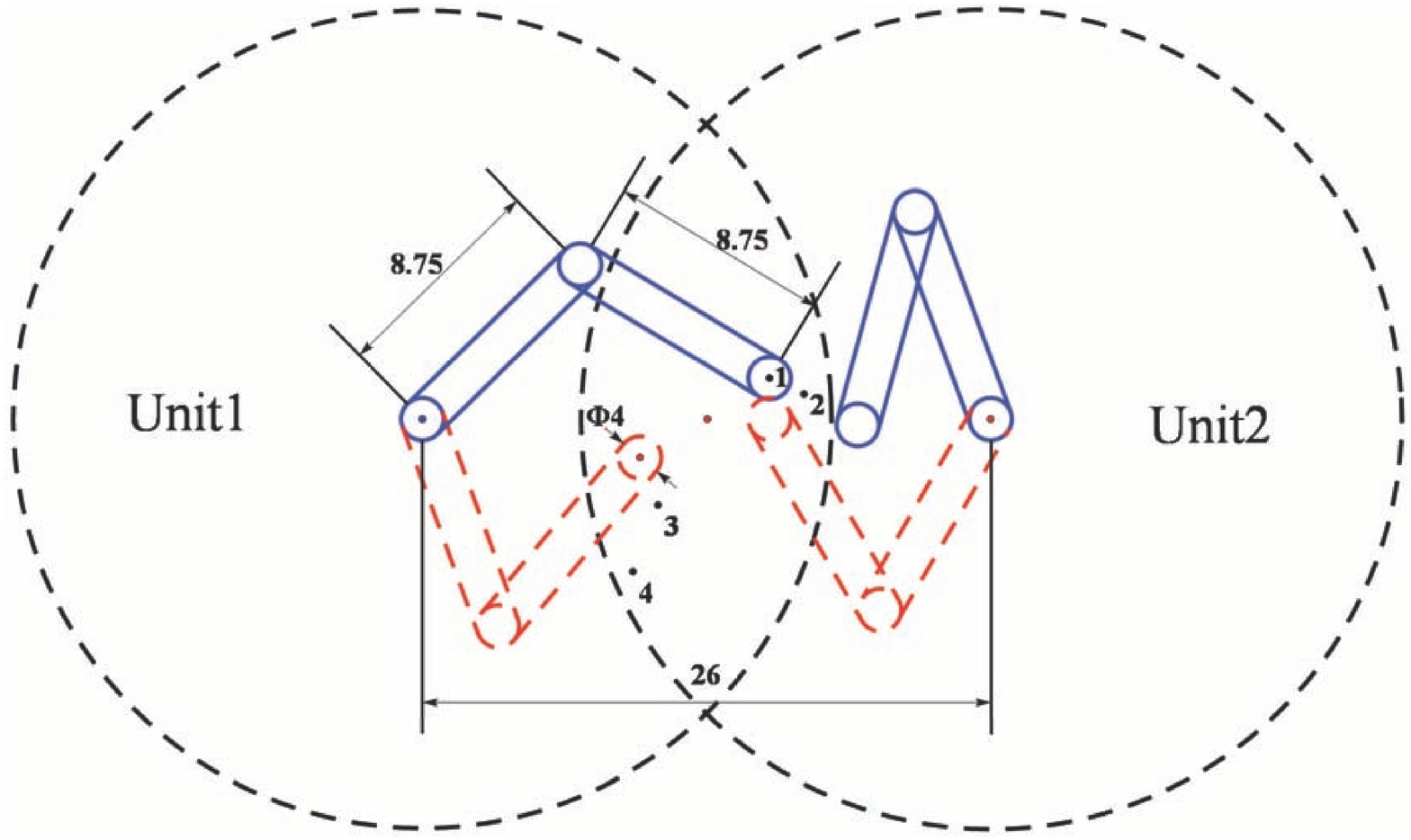}~~\includegraphics[height=5.0cm]{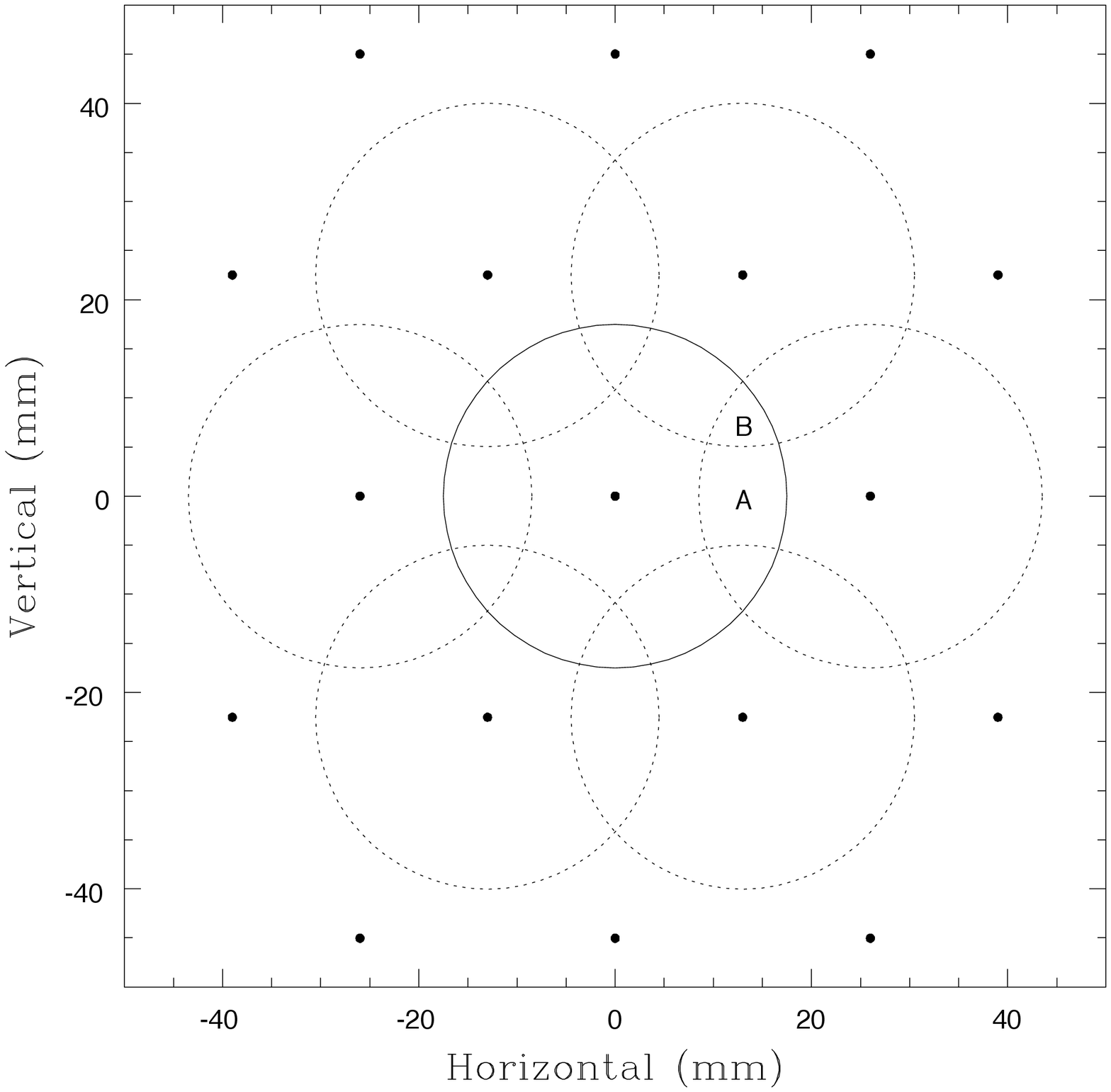}}
\caption{The left panel:  A plot showing the minimum separation of two targets on which two  fibers can be positioned. The right panel: The distribution of fibers on the focal surface. The distance between any two fiber centers is 26mm (or 4.47 arcmin). There are two
kinds of areas where fiber collision may happen, labeled ``A'' and
``B'' respectively. These areas actually gives the possibility to
measure objects in pairs (in ``A") or triples (in ``B'') at a single shot.
\label{fiberdist}}
\end{figure}

The fibers used by LAMOST have a scale of 3.3 arcseconds on the focal surface. The spectroscopic fiber centers are about 4.47 arcminutes
apart (the dots on the right panel of Figure~\ref{fiberdist}), and each fiber can be moved a maximum of 3 arcminutes from its central position (Xing et al. 1998).
The left panel in Figure~\ref{fiberdist} illustrates how the fibers are operated. Each fiber is driven a fiber robot made with two motors, which can place the fiber head to any point within a radius of 17.5 mm, with a forbidden range of 8mm around the mounting position. The right panel in Figure~\ref{fiberdist} shows how neighboring fibers can be placed. There are two possibilities of fiber collisions to be avoided, as demonstrated in the plot. Areas labeled ``A'' can be reached by two fibers, while regions marked  ``B'' are within reach of three fibers. These conditions actually define the closest pair or trio of targets that can be observed, and place fairly strong constraints on the uniformity of the targets over the field of view. For example, if an open cluster is 20' in diameter, we can place at most about 50 fibers on stars within the cluster's diameter, and those must be fairly uniformly distributed over the cluster area.  We can select fewer than 20 targets in the vicinity of a globular cluster with
tidal radius of 10'.  It will also be difficult to do completely filled surveys of any area
of the sky.  Many programs might benefit from combining observations to overcome fiber
spacing limitations.

The LAMOST facility will put 4000 fibers in 20 $\mathrm{deg}^2$ of sky
(200 fibers/$\mathrm{deg}^2$).  For comparison, the UKST places 100 fibers in 28 $\mathrm{deg}^2$ (5 fibers/$\mathrm{deg}^2$) and SDSS places 640 fibers in 7 sq. deg. (90 fibers/$\mathrm{deg}^2$).  Note, however, that the SEGUE survey observes two plates in each position on the sky, so the density of targets in SEGUE fields is about 180 fibers/$\mathrm{deg}^2$,
which is a close match to the fiber density in LAMOST.  Both UKST and SDSS
have much less stringent constraints on how far a fiber
can be moved from its nominal position on their focal planes.

\subsection{Wavelength Coverage and Spectral Resolutions}

Sixteen spectrographs are used in the system, each fed with the dispersed light from 250 fibers (Zhu et al. 2006, Zhu et al. 2010). 
The ``native'' resolution of the default LAMOST grating setup is $R\approx1000$, which yields spectral wavelength coverage of $3700 < \lambda < 9100$ \AA\. . The resolution can be increased by placing a slit of fixed width in front of the fiber heads (e.g., a slit of half the fiber width yields $R\approx2000$ while retaining the same wave length coverage). The survey will also include an $R=5000$ mode, which 
will yield two pieces
of the spectrum that are 350 \AA\ wide, one in the red and one in
the blue.  The blue wavelength coverage is centered around
$5300$ \AA\, to sample many metal lines, including the prominent Mg b (5175 \AA) triplet.  The red
segment covers the spectral range 8400-8750 \AA\., sampling the
CaII triplet, Fe I, Ti I, and other lines, which are ideal for measuring the RV
and $\mathrm{[Fe/H]}$.  This $R=5000$ mode wavelength coverage and dispersion is similar to that of the RAVE
experiment.  The $R=2000$ (half slit width) spectra will be similar in quality to the
longer exposure SEGUE spectra, with radial velocities and
metallicities determined to $\sim$7 km s$^{-1}$ and 0.3 dex, respectively.
The accuracies in measuring radial velocity and $\mathrm{[Fe/H]}$ at $R=5000$
are expected to be 1 km/s and 0.1 dex, respectively.

In the past 2 years, LAMOST has tested different resolutions by tuning the slit width, with the goal of having adequate resolution stellar spectra to accurately measure [Fe/H] (and possibly [$\alpha$/Fe]), without sacrificing too much light by placing a narrow slit in front of the fiber aperture. 
Considering all the possible science goals of the project, a consensus has been reached in the community to use only a fixed slit width of 2/3 the fiber diameter for the survey when R1000 gratings are mounted. This will give a spectral resolution of $R\approx1800$ around g band (similar to that of SDSS) instead of 1000 when no slit is used. This results in only about 22\% loss of light (compared to 40\% when a half width slit is used), while achieving a reasonable resolution. Whenever a slit will be inserted, the width will always be 2/3 for the survey. For the high resolution mode when R5000 gratings are used, no slit will be used in order to retain as many of the incoming photons as possible.

\subsection{A brief description of the data reduction procedure}

LAMOST data are processed by data pipelines written specifically for the LAMOST Spectral Survey (Zhao 2000). The data flow in the software chain is outlined here: raw data after observations will be transferred from the site to the data center at NAOC through a dedicated fiber connection on a daily basis. A cluster of 16 multi-processor workstations are used for 2D and 1D reduction pipelines. Tests during commissioning time have shown that the computing power is sufficient for the workload.

The LAMOST 2D pipeline follows the regular steps of the SDSS spectro2d pipeline \cite{spec2d}, but is different in many details. Twilight flat-fields taken each observing day are used to get the trace of each fiber spectrum. All the spectra from the same night are extracted using the same trace function, minor shifts are made to align individual frames, and then the spectra are flat-field corrected.
Extracted spectra are divided by the one dimensional flat-field.  
Hg/Cd and Ne/Ar arc lamp spectra are extracted to determine the dispersion function of each fiber, with strong sky emission lines also included to further fix the wavelength solution. 
About 20 sky fibers are allocated in each (250-fiber) spectrograph. After wavelength calibration, spectra from these sky fibers are combined into one super-sky spectrum. The super sky spectrum is scaled to best fit the sky spectrum in each fiber, then this scaled sky spectrum is subtracted from each fiber. Five F-type subdwarfs are observed in each spectrograph to provide flux calibration. Flux corrected spectra from different exposures taken on the same night are combined to improve the signal to noise ratio. The whole processing CPU time for one observation (three 30 minute exposures for all 4000 fibers) from tracing to flux calibration takes about 90 minutes. 

The 1D pipeline is similar to the SDSS SpecBS pipeline \cite{specbs}. The extracted spectra are first classified into 3 different categories (i.e., stars, galaxies, and QSOs) by a comparison of the observed spectra with templates. The radial velocity and redshift are then calculated.  After that, each object in each category is further processed
to determine the spectral subclass.
Galaxies are classified into subclasses AGN, STAR\_FORMING, STAR\-BURST, BROAD\-LINE and Normal, then the velocity dispersion of each galaxy is calculated by the synthetic stellar 
population method. Stars are classified into different spectral classes by matching to spectral templates.  The radial velocity of each stellar spectrum is determined by cross-correlation with the ELODIE \cite{ELODIE}
stellar spectral library. The final classification and redshift is written into the fits header of the spectrum. The whole procedure for 1D pipeline to reduce spectra of 
one observation (4000 fibers) takes about one CPU hour.

Since both the pipeline and the hardware need further refinement, each spectrum is examined by eye to ensure the quality. Those spectra with obvious problems are rejected.
After visual inspection, spectra of sufficient quality are released to the public (i.e., those who have been authorized for data access).

\begin{figure}
\centerline{\includegraphics[width=12cm]{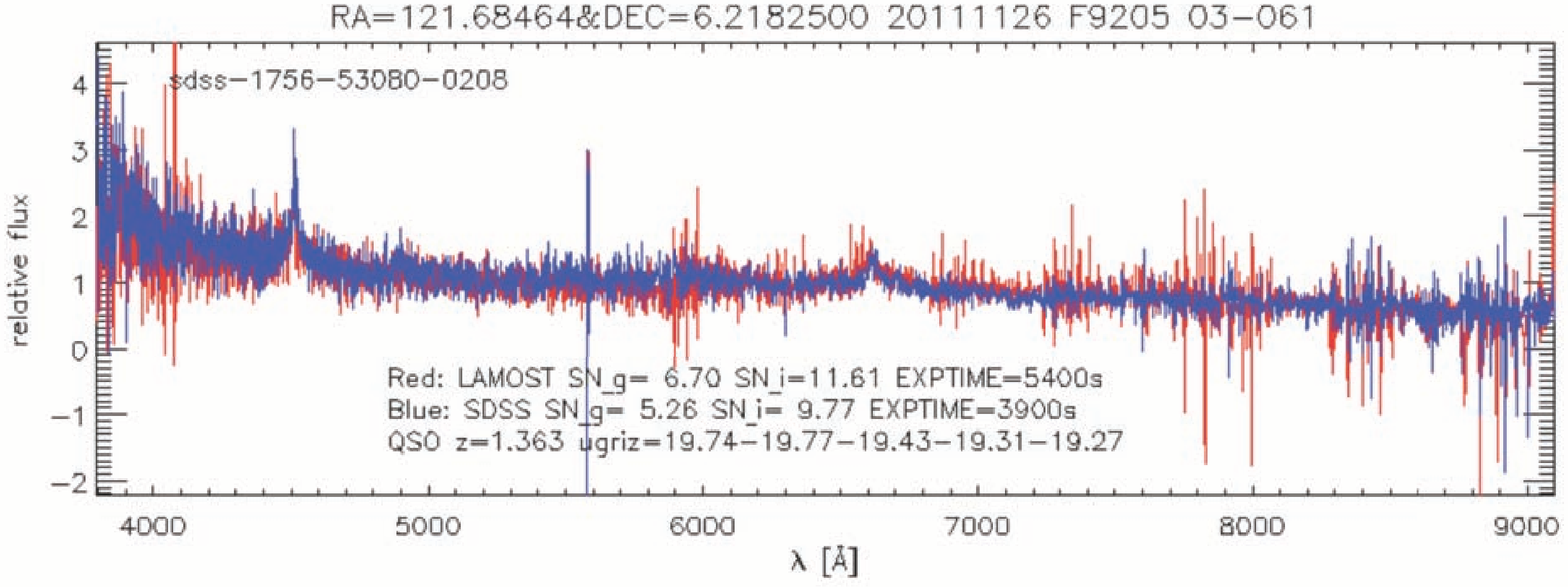}}
\centerline{\includegraphics[width=12cm]{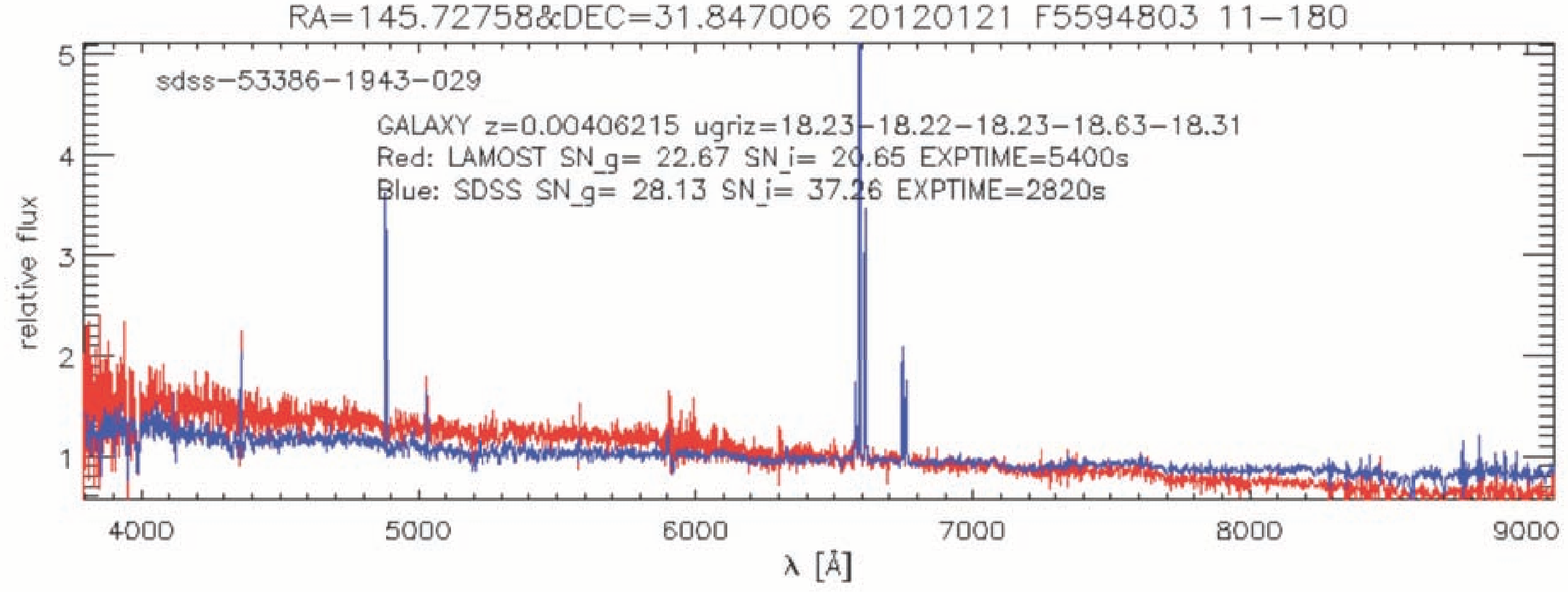}}
\centerline{\includegraphics[width=12cm]{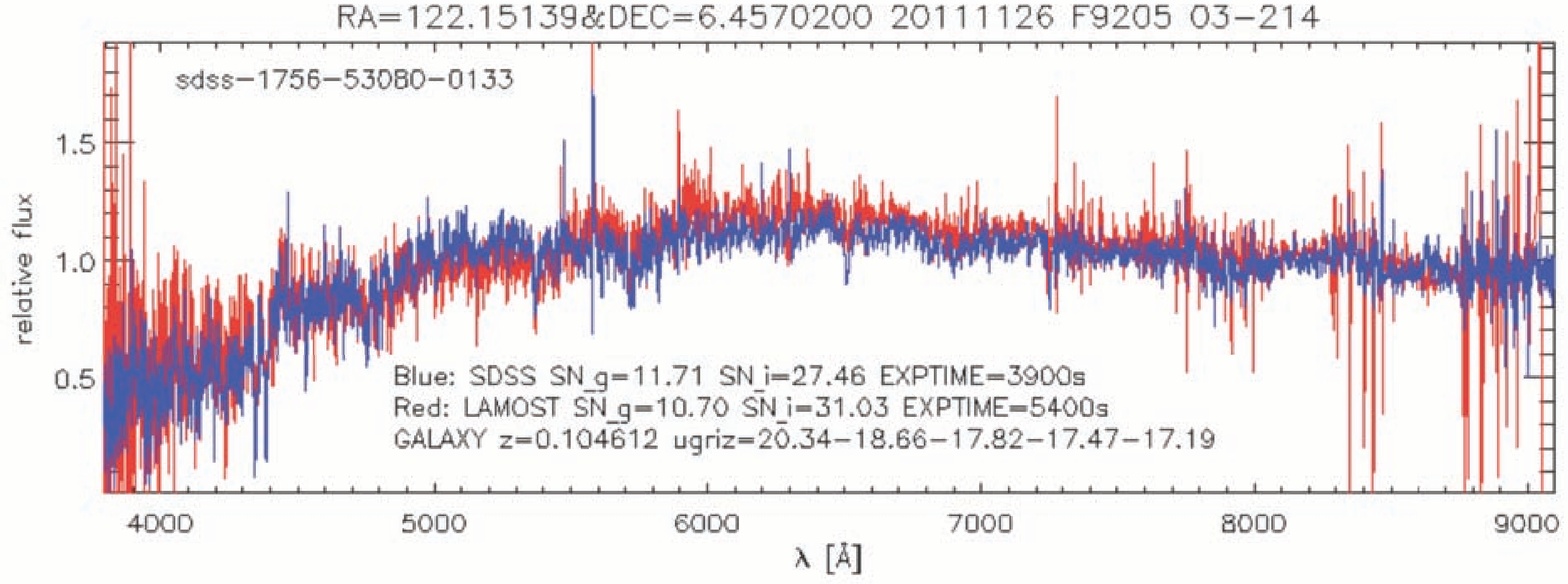}}
\centerline{\includegraphics[width=12cm]{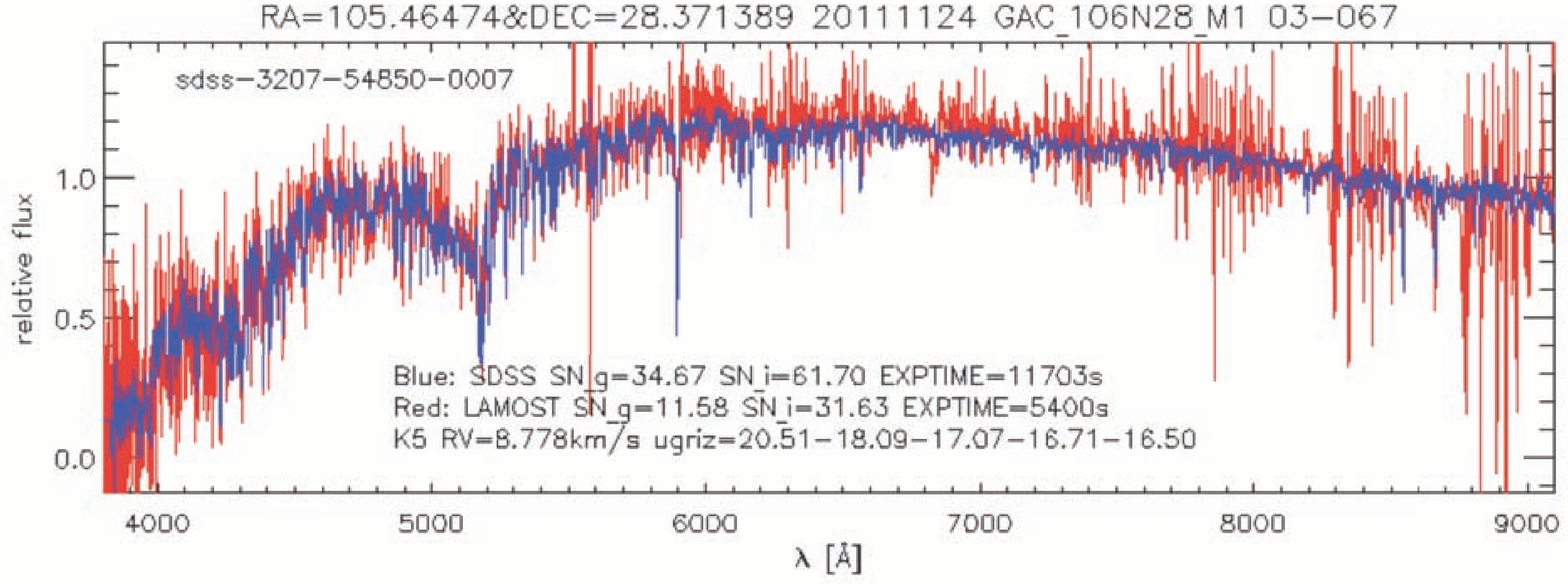}}
\caption{The LAMOST specta (red) of different types of objects are compared to the SDSS spectra (blue) for common objects, spectra are rescaled to similar exposure time and resolution.  The LAMOST sky subtraction and response function still needs
more work (note the O2 absorption bands at 6880\AA ~\&  7600\AA) but it can already achieve simlar $S/N$ as SDSS. top panel: a QSO; the second from top: a emission line galaxy; the second from bottom: a regular galaxy; bottom: a star. The magnitudes in SDSS system are indicated in each panel.
\label{spectrum}}
\end{figure}

With the data collected during the pilot survey, we are beginning to be able to assess the performance of LAMOST in survey operations. At its best, LAMOST can deliver data at a quality comparable to SDSS. Fig. \ref{spectrum} shows (from top to bottom) sample spectra of a $g=19.77$ quasar, a $g=18.22$ emission line galaxy, a $g=18.66$ normal galaxy, and a $g=18.09$ K5 star (all in red), together with their SDSS counterparts (in blue). The exposure times and $S/N$ at $i$ band are indicated in each panel. The sky subtraction and flatfield calibration of the LAMOST spectra is still improving, with the hope of reaching SDSS quality by the time the regular survey starts.

\section{The Pilot Survey}

Before conducting the regular LAMOST Spectral Survey, a test run of the system in survey mode is required in order to check the real instrumental performance and assess the feasibility of the science goals originally proposed for the survey in 2009 (the LEGUE portion is presented in this volume; Deng \& Newberg et al. 2012). Using the designed system parameters, and informed by commissioning tests, the science working groups of LEGUE and LEGAS came up with a plan for the LAMOST pilot survey. The pilot survey was launched on Oct 24, 2011. Here is a brief description:

\begin{itemize}
\item{{\bf Duration:} October 2011 - June 2012, including 9 full moon cycles. This basically covers all good observing season in a year at the site (see Yao et al. 2012 in this volume).}
\item{{\bf Time allocation:} Each moon cycle will be divided into 3 parts. dark nights: 5 days before and after the new moon respectively; bright nights: 5 days before and after full moon respectively; grey nights: the remaining nights in the cycle.  All time will be allocated to the survey, except the 3 nights attached to both ends of the dark nights, which will be used for technical purposes.}
\item{{\bf Observation plan: } In principle, dark nights will be used for faint objects (stars and extra-galactic sources). The extragalactic (LEGAS) sources require better observing conditions; therefore LEGAS sources (for the LEGAS footprint, see Figure 1 in Yang \& Carlin et al. 2012 in this volume) will have a higher priority whenever site conditions are optimal to reach $r=18$mag or better. On bright nights we will only target sources with any magnitude in \{g,r,i,z\} bands brighter than 16.5mag. When seeing and/or extinction become bad on dark nights, sources from the bright night catalogs will be targeted. The exposure time for dark night plates will be $3\times30$ minutes, and for bright plates exposures will be of order $3\times10$ minutes, depending on the distribution of the brightness of the sources.}
\item{{\bf Add-on projects}: LAMOST opens an opportunity for some special research projects (named ``add-on'' projects) that need only a small number of fibers by reserving less than 10\% fibers. This kind of observations can be proposed to the operation center at any time, and will be scheduled once approved by the Science Committee of LAMOST. Such proposals will subject to a number of constraints: the sources should be almost uniformly distributed in the sky; the number density should not be too high (lower than 10/degree$^2$); and should be within the magnitude range of the regular survey. During the pilot survey, a number of proposals have been received. One of them that meets the requirements and has been carried out is observing UV-excess targets, the science goal of which is to search for stellar mass black holes and neutron star binaries.}

\end{itemize}

Data collected in the Pilot survey will go through the same data processing system as the regular survey. However, the processed pilot survey data will be released to the community in a different way from that for the regular survey. The pilot survey data releases will be done monthly. Access to the released pilot survey data requires authorization by the operation center, which will be granted after the request of the interested individuals. In general, anyone who is interested in doing data analysis and is willing to help LAMOST by sending feedback to the operation center will be granted data access. A quick view of the data summary plots for each plate and a channel for downloading 1D extracted and calibrated survey spectra is available on our web site once the permission (by password) is given.

\section{The science plan for the survey}
\label{sect:sciplans}

\par
The both large aperture and large field of view advantage afforded by GSJT, combined with the high data rates made possible by 4000 fibers, provides us with an extraordinary opportunity to carry out the largest stellar and galactic survey ever conceived. The ability to observe millions of galaxies (LEGAS) and even larger numbers of stars in the Milky Way Galaxy (LEGUE) will open up new windows in near-field and far-field cosmology (Freeman \& Bland-Hawthorn 2002). In this volume, we will demonstrate LAMOST's potential for understanding the formation and evolution of our Galaxy.

The study of stars in the Milky Way galaxy is critical to understanding how galaxies form and evolve. Through study of galaxy formation, we test models of dark matter, gravitational collapse, hydro- dynamics of the gas, stellar formation and feedback (including properties of the first generation of stars and enrichment of the interstellar medium through supernova explosions). The Milky Way is the only galaxy we can study in enough detail that these models can be tested in three dimensions. Only recently have large photometric sky surveys, including the Sloan Digital Sky Survey (SDSS) and the 2 Micron All-Sky Survey (2MASS) made it possible to piece together the structure of the Milky Way star by star. Large spectroscopic surveys like the RAdial Velocity Experiment (RAVE), which targets only the brightest stars; the Sloan Extension for Galactic Understanding and Exploration (SEGUE), which very sparsely samples the sky; and the APO Galactic Evolution Experiment (APOGEE), which will observe bright red giant stars, are recently completed or currently in progress. However, there is an aching need for much larger, deeper, and denser spectroscopic surveys of Milky Way stars.
We plan to study the structure of the galactic halo (both the smooth component of the spheroid and the lumpy sub-structures) and disk components (including star-forming regions and open clusters). The revealed structure will inform our models of star formation, the formation history of the Galaxy, and the structure of the gravitational potential, including the central black hole and (sub)structure of the dark matter component. 

Within 4-5 years starting from the fall of 2012, LAMOST will observe at least 2.5 million stars in a contiguous area in the Galactic halo, and more than 7.5 million stars in the low galactic latitude areas around the plane. The spectra collected for such a huge sample of stars will provide a legacy that allow us to learn detailed information on stellar kinematics, chemical compositions well beyond SDSS/SEGUE.  Such a dataset will allow for studies of the Galaxy before Gaia era.  These science highlights and the procedures that have been implemented to realize these goals are discussed in detail in this volume.

The pilot survey, aimed at learning about the system performance and testing the science goals started in October last year, and is still ongoing. In this volume, we give the public an overview of the LEGUE project, and the LEGUE portion of the pilot survey in particular.  In consideration of the science goals and the resources available, LEGUE is divided into 3 parts in sky coverage, namely Galactic spheroid, the disk and the Galactic anticenter (GAC).  The science plans and technical designs for the pilot LEGUE survey, except that of GAC, will be presented in this special issue. The GAC part of LEGUE requires independent photometry for target selection; the photometric survey has been done, and the science plan for GAC will appear at a later time (Liu et al., in preparation). 

LEGAS is the other main component of the survey that requires better system performance.  The LEGAS plan will be addressed by a separate, future papers.

The current volume includes the following:

\begin{enumerate}
\item LEGUE science plan (Deng et al. 2012), 
\item The general target selection algorithm for LEGUE (Carlin et al. 2012), 
\item The design of LAMOST pilot survey in bright nights (Zhang et al. 2012)
\item The design of LAMOST pilot survey in dark nights (Yang et al. 2012)
\item The design of LAMOST pilot survey for the disk of the Galaxy (Chen et al. 2012)
\item The site conditions for LAMOST spectral survey (Yao et al. 2012)
\end{enumerate}

The purpose of the current volume is to help astronomers to understand the grand design of the instruments and the survey. More details on the survey and data reduction are forthcoming along with data releases by the operation center.

\section{Policies}

LAMOST is a national key scientific research facility.  The Chinese national funding agencies provided full coverage of construction and operations for the facility and the science missions, with a hope that such a tremendous effort will be rewarded by scientific output. Therefore, the entire Chinese astronomical community will be automatically involved in research based on the survey data. In order to maximize science output of the facility, international collaboration has been highly encouraged since the beginning of the project. The project has benefitted greatly from contributions of the international community, including reviews of the project and science planning for the survey, and with expertise provided individuals and groups. The National Astronomical Observatories, Chinese Academy of Sciences is responsible for running the facility and its science operations. The operation center, assigned by the Chinese Academy of Sciences, has published the policies for data access and publications based on the survey database. For those who will be interested in using LAMOST spectral survey data for their research, there are a number of ways to participate, as outlined by the data policy and publication policy. Please visit our website for more information: http://lamost.org.


LAMOST data will be classified into 3 main types: raw data (target, flat and calibration images), 1D spectra (raw data after processing and extraction by the 2D pipeline) and object catalogs (parameters produced from the 1D pipeline processing). Auxiliary data, including weather condition sensors and guiding camera images, etc., will also be archived for further data quality assessments.  LAMOST data releases will contain 1D spectra and object catalogs. Data collected for the survey will have an 18 month proprietary period after data taking for research projects organized by the operation center.  Data will be made publicly available after this proprietary period. For a detailed description of the LAMOST data, please check the official data policy on the LAMOST homepage (http://www.lamost.org/policies/data\_policity.html). 
Applications to use the publicly released LAMOST survey data for all purposes are highly encouraged; rules for publishing scientific results derived from LAMOST data are described in the publication policy (http://www.lamost.org/policies/paper\_policy.html).

\acknowledgement

 L.D. thanks NSFC for support from grants Nos. 10573022, 10973015 and 11061120454. It is acknowledged that Heidi J. Newberg from RPI and the her team PLUS (Participating LAMOST, US) have made substantial contributions in designing the survey under the support of NSF through grant AST-09-37523. We thank the referee, Joss Bland-hawthorn for constructive comments and suggestions.

\end{document}